\documentclass[
reprint,
amsmath,amssymb,
aps,
pra,
]{revtex4-2}

\usepackage{graphicx}
\usepackage{dcolumn}
\usepackage{bm}
\usepackage{hyperref}
\usepackage{float}
\usepackage[table]{xcolor}
\usepackage{array}
\usepackage{makecell}
\usepackage{siunitx}
\usepackage[version=4]{mhchem}
\usepackage[utf8]{inputenc}
\usepackage{mathtools}

\DeclareUnicodeCharacter{03C9}{\ensuremath{\omega}}
\DeclareUnicodeCharacter{2009}{\,}

\DeclareSIUnit{\dBm}{dBm}
\newcommand{\Ej}{\ensuremath{E_\text{J}}}

\newcommand{\omegaJ}{\ensuremath{\omega_\text{dc}}}
\newcommand{\phiJ}{\ensuremath{\phi_\text{dc}}}
\newcommand{\fJVS}{\ensuremath{f_\text{JVS}}}
\newcommand{\PJVS}{\ensuremath{P_\text{JVS}}}
\newcommand{\fsignal}{\ensuremath{f_\text{S}}}

\newcommand{\fdc}{\ensuremath{f_\text{dc}}}
\newcommand{\Vdc}{\ensuremath{V_\text{dc}}}
\newcommand{\SIref}[1]{Sup.\ Mat.\ \ref{#1}}

\newcommand{\ain}{a_\text{in}}
\newcommand{\aout}{a_\text{out}}

\newcommand{\dets}{\eta_\text{s}}
\newcommand{\deti}{\eta_\text{i}}

\begin{document}
\title{Phase-stable voltage bias for Josephson photonics devices}
\author{Amir Hosein Esmaeili, Naveen Nehra, Alexandre Paquette, Mona Arabmohammadi, Baptiste Monge, Alexandre Rogalle, François Cyrenne-Bergeron, Yannick Lapointe, Nicolas Lavoie, Nicolas Bourlet, Max Hofheinz}

\affiliation{Institut Quantique, Université de Sherbrooke, Sherbrooke, Quebec J1K 2R1, Canada}

\date{\today}

\begin{abstract}

Parametric interactions are foundational to superconducting quantum technologies, yet conventional microwave-driven pumping introduces parasitic Kerr nonlinearities and higher-order harmonics that limit device performance. Josephson Photonics (JP) avoids these parasitics by utilizing dc-biased junctions but has remained constrained by high phase noise and the absence of a stable phase reference. Here, we overcome this limitation by integrating a Josephson voltage standard (JVS) to establish a noise-tolerant phase reference, and demonstrate that this reference is coherently transferred to an inelastic Cooper-pair tunneling amplifier (ICTA) via the superconducting order parameter.The resulting low-phase-noise architecture yields a 14-dB enhancement in averaged gain and better quantum-limited noise performance. Critically, the phase reference enables the first observation of phase-sensitive gain and sqeezing in a dc-biased amplifier. By reconciling clean, Kerr-free nonlinearities with phase-coherent drive, our architecture establishes a robust platform for high-purity parametric processes in superconducting circuits.

\end{abstract}

\maketitle

Parametric interactions are essential building blocks in superconducting quantum technologies. They enable quantum-limited amplification \cite{bergeal2010phase,parker2022degenerate}, single-photon detection \cite{petrovnin2024microwave}, high-fidelity entangling gates between transmon qubits in superconducting quantum processors \cite{wu2021strong}, and the stabilization of bosonic error-correction codes \cite{caldwell2018parametrically,ma2021quantum}. These interactions typically rely on microwave pumps with well-defined phase to achieve precise control and high quantum efficiencies; however, this approach introduces intrinsic parasitic effects, including higher-order pump harmonics and Kerr nonlinearities~\cite{boutin2017effect}. These parasitics manifest as detrimental phenomena such as shock-wave formation in traveling-wave parametric amplifiers (TWPAs) \cite{landauer1960shock}, limited saturation power in quantum-limited amplifiers \cite{macklin2015near}, and degraded gate fidelity or size limits in bosonic codes \cite{putterman2025preserving}.

In contrast, Josephson Photonics (JP) devices~\cite{jebari2018near,rolland2019antibunched,peugeot2021generating,menard2022emission,albert2024microwave,nehra2026broadband,paradina2026engineering} generate the same parametric interaction terms using dc-biasing Josephson junctions. The distinction between these pumping schemes arises from how the parametric drive enters the junction Hamiltonian \cite{aissaoui2024cat}:
$$
\begin{aligned}
	U_{\mathrm{\mu w}} &= -E_J \cos \bigl( \varepsilon_p \cos(\omega_p t) + \phi_\text{p} + \varphi\bigr), \\
	U_{\mathrm{dc}} &= -E_J \cos \bigl( \omegaJ t + \phiJ + \varphi\bigr),
\end{aligned}
$$
where $ U_{\mathrm{\mu w}} $ and $ U_{\mathrm{dc}} $ are the potential energies of the microwave-driven and dc-biased Josephson junctions, respectively and $\varphi$ the nonclassical part of the Jospehson junction phase through which the junction couples to the rest of the circuit. The parameters $ \varepsilon_p $, $ \omega_p $, and $ \phi_p $ denote the amplitude, angular frequency, and phase of the microwave pump tone. The dc voltage $ V_{\mathrm{dc}} $ gives rise to Josephson oscillations at angular frequency $ \omegaJ = 2\pi f_{\mathrm{dc}} $, where $\fdc = 2eV_{\mathrm{dc}}/h $ is the corresponding Josephson frequency, and $ \phiJ $ denotes the Josephson phase.
The simpler temporal dependence of \( U_{\mathrm{dc}} \) directly suppresses the generation of higher-order pump harmonics and Kerr nonlinearities, offering a significant advantage over microwave-driven systems.
However, bringing to use this ``cleaner'' nonlinearity in JP devices is challenging: The Josephson frequency in JP devices has a linewidth typically 6 to 7 orders of magnitude larger than a microwave source~\cite{domenico2010, martel2025influence}, and does not have a well-defined phase reference, leaving \(\phiJ\) ill-defined, precluding phase coherent operation~\cite{paradina2026engineering, aissaoui2024cat,leppakangas2014input,peugeot2021generating}.
 
These challenges are exemplified by the inelastic Cooper-pair tunneling amplifier (ICTA)~\cite{jebari2018near,martel2025influence,nehra2026broadband}. The ICTA is based on parametric down-conversion of $\omega_\text{dc}$ into a signal and idler photon. While the ICTA benefits from a wide bandwidth and near-quantum-limited noise using a minimal circuit of two junctions~\cite{nehra2026broadband}, its performance is constrained by phase noise. When the voltage-noise-induced linewidth exceeds the device bandwidth, voltage noise is transferred to phase noise in amplified signal,  limiting gain and increasing noise~\cite{martel2025influence}. Furthermore, the lack of a well-defined Josephson phase \(\phiJ\)---the time integral of the bias voltage---prevents phase-sensitive operation in degenerate configuration \cite{jebari2018near}.

In principle, injection locking a Josephson junction with a strong microwave tone at the Josephson frequency can mitigate these limitations by suppressing dc voltage noise and establishing a well-defined junction phase \cite{danner2021injectionlocking, cassidy2017}. However, robust injection locking requires Josephson junctions with high Josephson energies and has so far prevented experimental implementation of multi-photon parametric interactions in JP \cite{danner2021injectionlocking, aissaoui2024cat, hohe2025quantum, danner2025stabilizing}. This difficulty arises because multi-photon processes typically require high-impedance environments and smaller Josephson energies, rendering injection locking more fragile and highly sensitive to noise. Consequently, despite their intrinsically cleaner nonlinearities, JP devices so far have not been used for applications requiring phase-coherent operation, such as phase-sensitive amplification and cat-code stabilization~\cite{aissaoui2024cat,leppakangas2014input,peugeot2021generating, paradina2026engineering}.

\begin{figure}[t]
	\raggedright
	\includegraphics[width=\columnwidth]{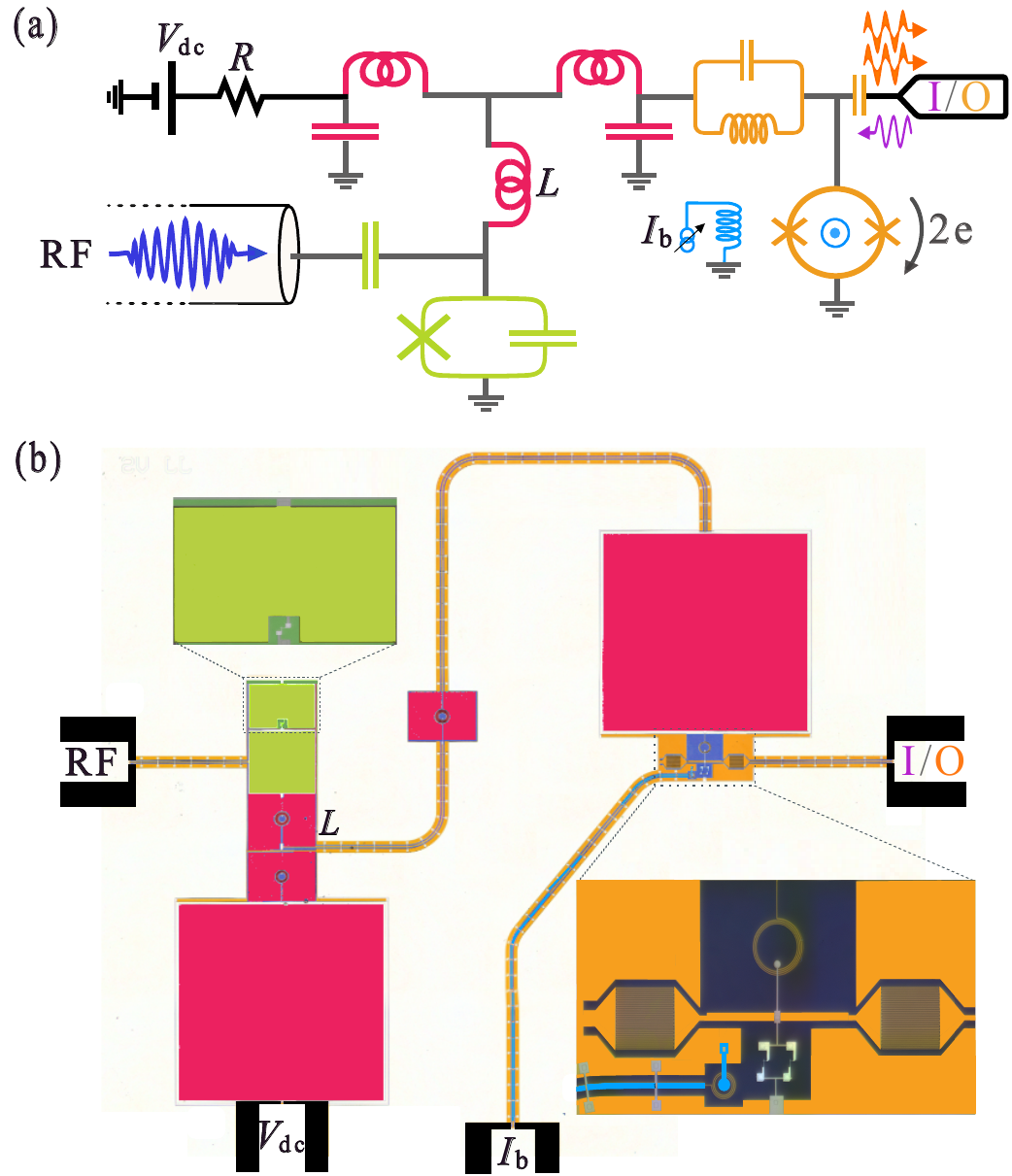}
	\caption{\label{fig:fig1} \textbf{Integrated JVS-biased ICTA device. }
	\textbf{(a)} Circuit schematic of the device containing the JVS, ICTA, and superconducting low-pass filter. The JVS consists of a Josephson junction in parallel with a capacitor (shown in green) it is dc biased with a voltage $V_\text{dc}$ through a source impedance $R=\SI{5}{\ohm}$ and an RF tone is capacitively coupled. The ICTA comprises a single lumped-element resonator grounded by a SQUID (shown in orange) whose magnetic flux is controlled by an on-chip flux-bias line (shown in blue). Input signals are sent through a capacitively coupled I/O port and get amplified in reflection. The two devices are connected via a low-pass filter  (shown in red), isolating the two devices at microwave frequencies while providing a superconducting dc connection.
	\textbf{(b)} Optical microscope image of the fabricated device on a \(5 \times 5\,\mathrm{mm}^2\) chip, false-color coded to match the elements in the schematic.
}
\end{figure}

In this work, we overcome these limitations by generating a phase reference from an RF-driven Josephson voltage standard (JVS), based on Shapiro steps described by classical dynamics~\cite{rufenacht2018impact,tang201210, smirr25tunable}, and employ it to stabilize a voltage-biased parametric interaction in an ICTA \cite{jebari2018near,martel2025influence,nehra2026broadband}.
This hybrid approach decouples the requirements of the two systems: a JVS with large junctions provides a robust phase reference, while the ICTA employs small junctions optimized for parametric conversion. We demonstrate that the JVS phase reference is transferred to the ICTA through the superconducting order parameter. We observe enhanced gain and noise performance and, most importantly, phase-sensitive operation of the ICTA with generation of vacuum-noise squeezing. To implement this phase-coherent architecture, we couple the JVS to the ICTA through a superconducting low-pass filter, providing a superconducting dc link while suppressing microwave-frequency interactions between the two devices. The complete circuit is fabricated on a single $  5 \times 5\,\mathrm{mm}^2  $ chip (see Fig.~\ref{fig:fig1}, \SIref{sec:Fabrication}).

JVSs are typically engineered for metrology and operate on a \SI{10}{\volt} range at $\SI{4}{\kelvin}$ \cite{hamilton2000josephson,kohlmann2003josephson}. They require rigorous optimization of junction parameters to ensure well-defined Shapiro steps and prevent the system from entering a chaotic state \cite{kautz1996noise}.
JVSs have also been demonstrated for quantum devices operating at millikelvin temperatures \cite{smirr25tunable} where they must be optimized differently: because microvolt-level voltages are required, only a single junction is needed; and because of the much lower cooling power available at millikelvin temperatures, the JVS must operate at lower RF power and use capacitively shunted junctions to minimize on-chip dissipation. 

In our device, the JVS (green in Fig.~\ref{fig:fig1}(a,b)), consists of a Josephson junction with area of \(0.5\,\si{\micro\meter\squared}\) and a critical current \(I_\text{C,JVS} \approx \SI{3}{\micro\ampere}\). With a plasma frequency of approximately \SI{1.5}{\giga\hertz}, this junction facilitates quantized voltage spikes across the \SIrange{4}{50}{\micro\volt} range required for biasing the ICTA.
For instance, an RF drive at \SI{12}{\giga\hertz} generates spikes of $n \times \SI{24}{\micro\volt}$. By simulation and then experiment, we find the first spike ($n=1$) remains robust for RF powers between \SI{-65}{\dBm} and \SI{-40}{\dBm}, yielding a maximal critical current of \SI{1.5}{\micro\ampere}. The height of this spike defines the locking range over which the JVS voltage is stabilized; increasing the RF drive power broadens this range, enhancing the device's resilience to voltage bias fluctuations (see \SIref{sec:Josephson Voltage Standard Design} and \SIref{sec:DC Measurement Setup} for design and measurement, respectively).

The JVS is coupled to the ICTA through an on-chip low-pass filter (red in Fig.~\ref{fig:fig1}(a,b)) to limit RF pump leakage from the JVS into the ICTA, while maintaining a continuous dc superconducting link to provide a phase-stable dc voltage via the superconducting order parameter. The filter exhibits approximately $\SI{50}{\decibel}$ stop-band rejection, except for a sharp parasitic resonance near \(\SI{10}{\giga\hertz}\) (see \SIref{sec:Filter Design}).

The ICTA (orange in Fig.~\ref{fig:fig1}(a,b)) comprises a lumped-element \(LC\) resonator at \SI{6}{\giga\hertz} terminated by a superconducting quantum interference device (SQUID) with junction areas of \SI{0.02}{\micro\meter\squared}. The voltage bias is applied to the SQUID through the resonator inductor, while an additional capacitor couples the mode to the I/O (input--output) transmission line. The coupled resonator has a linewidth \(\Gamma/2\pi \simeq \SI{1.2}{\giga\hertz}\), setting the amplifier's gain--bandwidth product (see \SIref{sec:Inelastic Cooper-pair Tunneling Amplifier Design}).

\begin{figure}[b]
	\raggedright
	\includegraphics[width=\columnwidth]{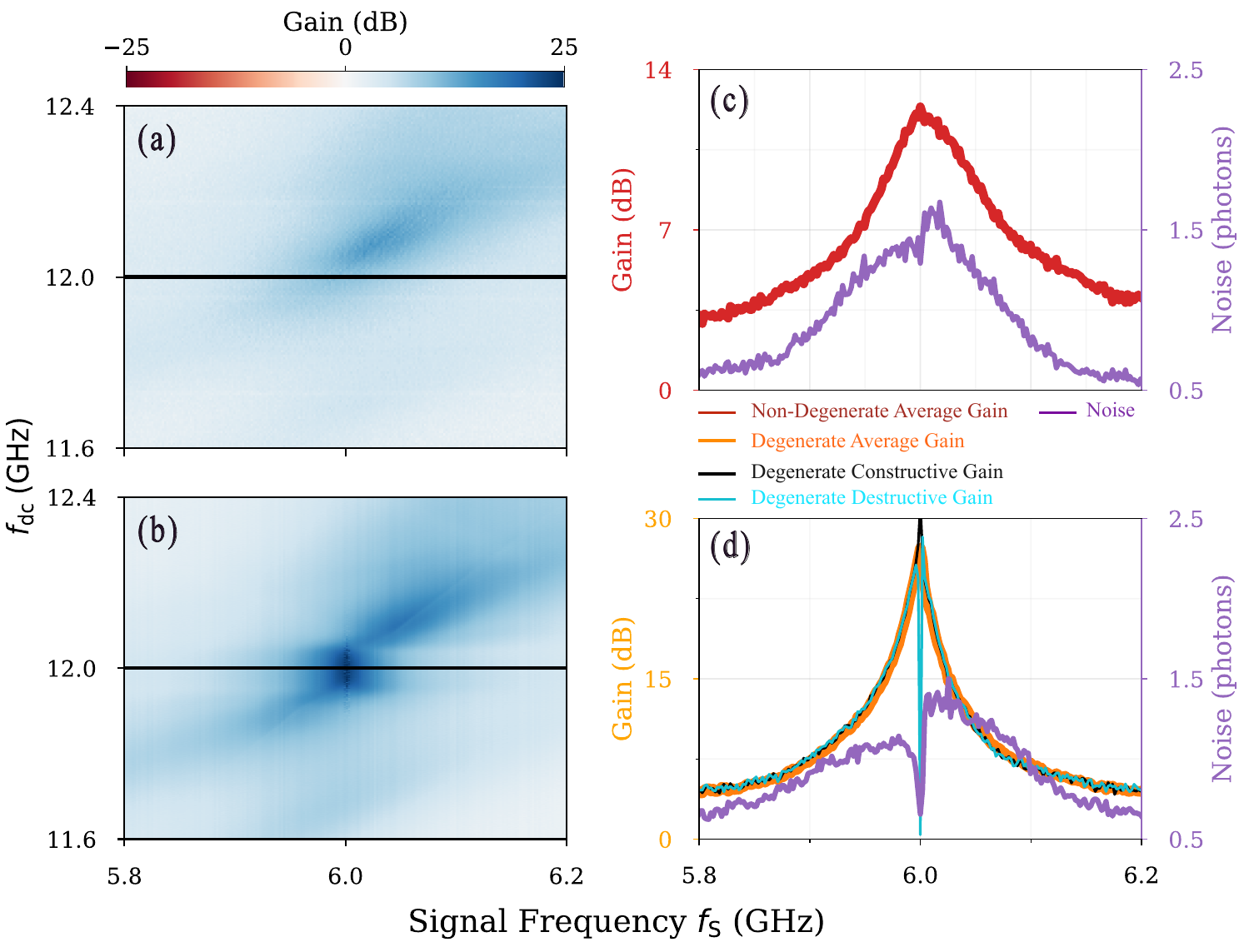}
	\caption{\label{fig:fig2}\textbf{Comparison of ICTA without and with JVS stabilization.}
		All data were acquired at a constant ICTA critical current $I_c \approx \SI{80}{\nano\ampere}$.
		\textbf{(a)} Two-dimensional gain map of the unstabilized ICTA (JVS RF tone off) as a function of signal frequency ($\fsignal$) and voltage bias expressed as Josephson frequency ($\fdc$); gain is maximal near the $LC$ resonance ($f_0=\SI{6}{\giga\hertz}$). 
		\textbf{(b)} Corresponding two-dimensional gain map with the JVS activated (JVS RF tone $\fJVS=\SI{12}{\giga\hertz}$, $\PJVS=\SI{-65}{dBm}$), locking the actual Josephson frequency at $\SI{12}{\giga\hertz}$ over a $\pm\SI{50}{\mega\hertz}$ bias range.
		\textbf{(c)} Cut of panel (a) at $\fdc = \SI{12}{\giga\hertz}$ showing gain of up to $\SI{13}{\dB}$ with a bandwidth of $\sim\SI{80}{\mega\hertz}$ (red curve) and an input-referred noise of $\sim 1.5$ photons (purple curve), no particular feature is observed at degeneracy $\fsignal = \fdc/2$ .
		\textbf{(d)} Cut of panel (b) at $\fdc = \fJVS = \SI{12}{\giga\hertz}$ showing average gain of up to $\sim\SI{27}{\dB}$ (orange curve). Non-averaged traces (gray and blue) show strong phase dependance of gain. The non-averaged input-referred noise (purple curve) is close to the quantum limit of 1 photon near degeneracy and $0.5$ photons at the degeneracy, corresponding to the black gain curve in panel (d).
	}
\end{figure}

To evaluate the impact of the Josephson voltage standard on amplifier performance, we compare the ICTA gain and noise characteristics with the JVS in both the inactive and active states (Fig.~\ref{fig:fig2}). Measurements were performed at a constant ICTA critical current ($I_c \approx \SI{80}{\nano\ampere}$) and a signal power of $\SI{-120}{\dBm}$.

With the JVS inactive, Fig.~\ref{fig:fig2}(a), the maximum gain is observed near the designed resonator frequency ($f_0=\SI{6}{\giga\hertz}$) and follows the expected degenerate parametric relation $\fsignal \simeq \fdc/2$. However, no particular feature signalling phase-dependent amplification is observed at exact degeneracy $\fsignal = \fdc/2$. This is expected when the phase \(\phiJ\) is sufficiently noisy to drift by more than $2\pi$ during the measurement time (see \SIref{sec:Phase-sensitive and phase-preserving regimes of the single cavity ICTA}). As a result, the squeezing angle is effectively randomized and degenerate parametric gain tends to non-degenerate gain. Under these conditions, the amplifier yields a maximum gain of $\SI{13}{\dB}$ across a bandwidth of approximately $\SI{80}{\mega\hertz}$ (red curve in Fig.~\ref{fig:fig2}(c)) and an input-referred noise of $\sim 1.5$ photons (purple curve).

Activating the JVS, by applying an RF tone at \(\fJVS=\SI{12}{\giga\hertz}\) with power $\PJVS=\SI{-65}{dBm}$), dramatically changes the device response, as shown in Fig.~\ref{fig:fig2}(b). Across the first Shapiro step--spanning approximately $\SI{11.95}{\giga\hertz}$ to $\SI{12.05}{\giga\hertz}$--the signal frequency corresponding to the gain maximum becomes independent of the applied voltage $\fdc$, indicating that the JVS rigidly locks the Josephson frequency to $\fJVS$. Prominent degenerate features then emerge at $\fsignal = \fJVS/2 = \SI{6}{\giga\hertz}$, indicating that the squeezing angle is stabilized by the JVS. As shown in Fig.~\ref{fig:fig2}(d), the stabilized gain (averaged over 10 traces with different squeezing angles) increases significantly to $\SI{27}{\dB}$ (orange curve), while the bandwidth contracts to below $\SI{5}{\mega\hertz}$. This observation is in excellent agreement with the gain--bandwidth trade-off: by suppressing the voltage-noise-induced linewidth, the ICTA reaches higher gain within a narrower bandwidth~\cite{martel2025influence}.

The non-averaged gain traces (black and blue curves of Fig.~\ref{fig:fig2}(d)) reveal sharp peaks and dips at exact degeneracy ($\fsignal = \fJVS/2$), depending on the squeezing angle. 

The stabilization of the Josephson phase upon activating the JVS also reduces the input-referred noise, decreasing it from $\sim 1.5$ photons shown in  Fig.~\ref{fig:fig2}(c)) to an average of $\sim 1$ photon. Here, we further show a non-averaged noise trace (purple curve in Fig.~\ref{fig:fig2}(d)) that approaches $0.5$ photons at the degeneracy point, coinciding with the constructive-gain condition observed in the black gain trace. These results demonstrate that the JVS suppresses Josephson-frequency phase noise, enabling a $\SI{14}{dB}$ gain enhancement and phase-sensitive operation of the ICTA.


\begin{figure}[t]
	\raggedright
	\includegraphics[width=0.9\columnwidth]{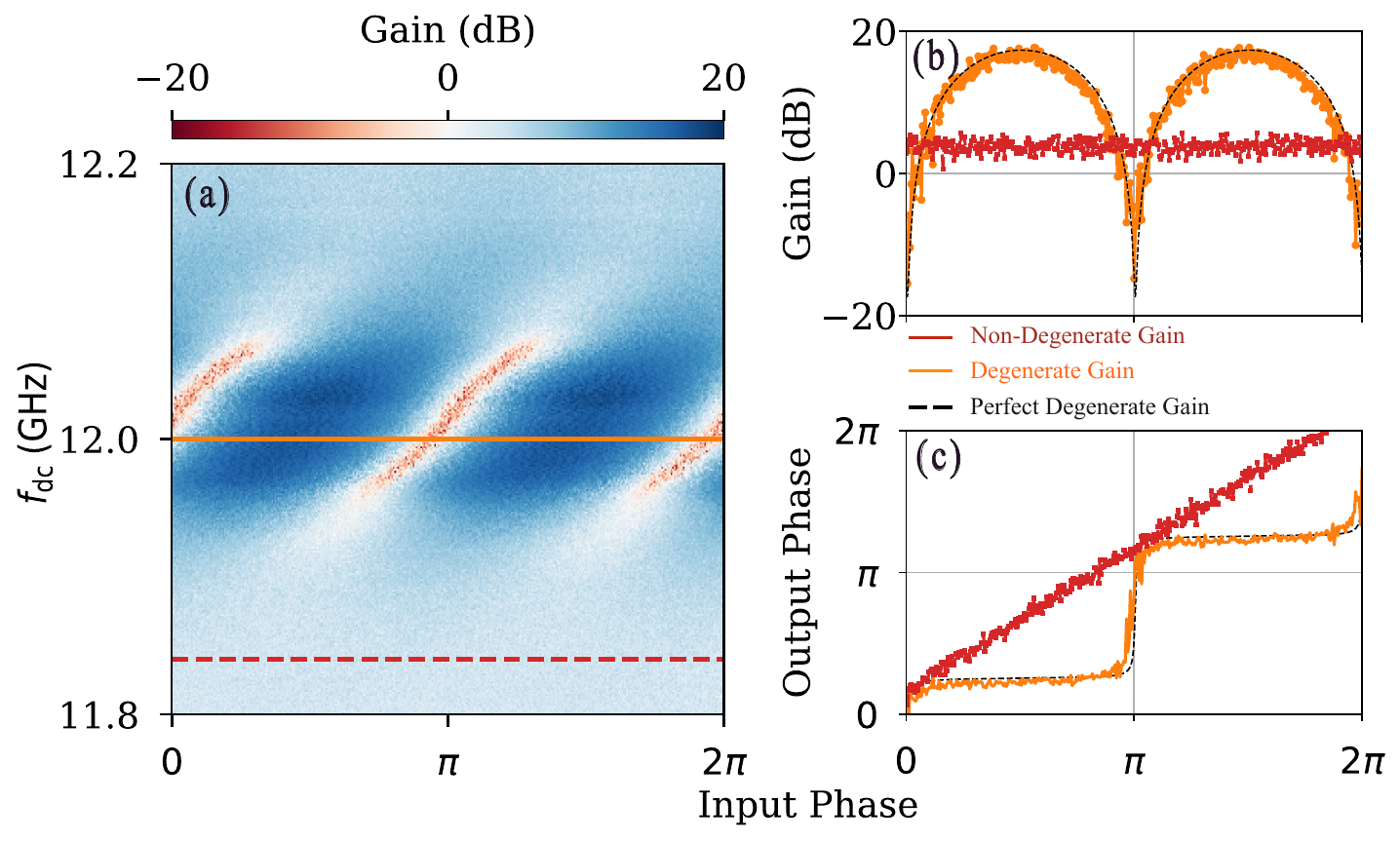}
	\caption{\label{fig:fig3} \textbf{Phase-dependence of gain.}
		\textbf{(a)} Gain map as a function of input signal phase and bias ($\fdc$). The map clearly delineates the transition from the phase-preserving regime to the phase-sensitive regime within the JVS locking range ($\SI{11.95}{\giga\hertz}$ to $\SI{12.05}{\giga\hertz}$).
		\textbf{(b)} One-dimensional gain cuts in the locking range ($\fdc \approx \fJVS = \SI{12}{\giga\hertz}$, orange) and outside the locking range ($\fdc = \SI{11.85}{\giga\hertz}$, red). The gain in the locking range shows the characteristic $\pi$-periodicity of a degenerate parametric amplifier, with an amplification range of $\pm\SI{17}{\dB}$, whereas the unlocked response remains phase-independent at $\SI{7}{\dB}$.
		\textbf{(c)} Output phase response corresponding to the measurements in panel (b). In the phase-sensitive regime, the phase evolution is step-like due to the quadrature-dependent gain, while the phase-preserving regime shows a strictly linear variation. Dashed lines in (b) and (c) represent perfect degenerate parametric amplifier magnitude and phase response with squeezing parameter $r=2$ (see \SIref{sec:Phase-sensitive and phase-preserving regimes of the single cavity ICTA}).
	}
\end{figure}

To further explore the phase coherence afforded by the JVS, in Fig.~\ref{fig:fig3} we examine the signal phase response at exact degeneracy ($\fsignal = \fJVS/2 = \SI{6}{\giga\hertz}$). For these measurements, a slightly reduced critical current ($I_c \approx \SI{65}{\nano\ampere}$) was used to clearly resolve the sharp transition between phase-preserving and phase-sensitive regimes; a detailed analysis of the higher-gain measurements is provided in \SIref{sec:Saturation in the degenerate regime}.

The two-dimensional magnitude gain map in Fig.~\ref{fig:fig3}(a) reveals two distinct regimes. Within the JVS locking range—spanning approximately $\SI{11.95}{\giga\hertz} \leq \fdc \leq \SI{12.05}{\giga\hertz}$—the gain exhibits the expected $\pi$-periodic input-phase dependence characteristic of the phase-sensitive regime. Outside this stabilization window, the gain becomes essentially phase-independent, characteristic of the phase-preserving regime. This contrast is explicitly quantified in the one-dimensional cuts shown in Fig.~\ref{fig:fig3}(b) and (c): while the non-degenerate gain remains constant at $\sim\SI{7}{\dB}$ for all input
phases (red curves), the degenerate response (orange curves) spans $\pm\SI{17}{\dB}$
and closely matches the expected profile of a perfect degenerate parametric amplifier
(dashed black curves, see \SIref{sec:Phase-sensitive and
phase-preserving regimes of the single cavity ICTA}).

\begin{figure}[t]
	\includegraphics[width =1\columnwidth]{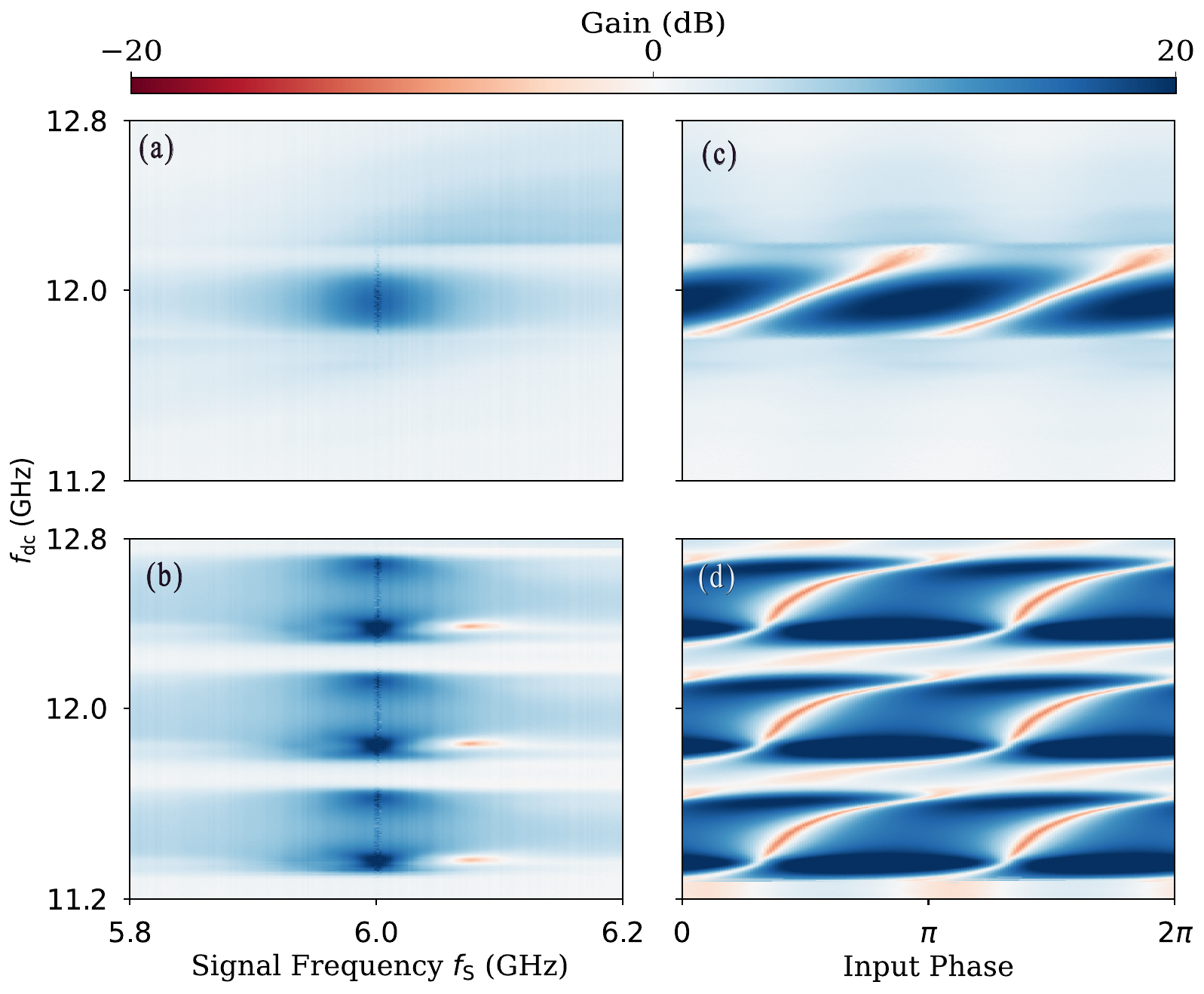}
	\caption{\label{fig:fig4}\textbf{Gain modulation within the JVS locking range.}
		\textbf{(a, b)} Magnitude gain of the ICTA as a function of bias $\fdc$ and signal frequency $\fsignal$ for two distinct JVS RF drive powers. At $\PJVS = \SI{-55}{\dBm}$ (a), a single gain-modulation period is observed across the locking range. Increasing the RF power (b) to $\SI{-43}{\dBm}$ broadens the locking range, revealing three distinct periods. The observed frequency periodicity $\Delta \fdc = \SI{0.5}{\giga\hertz}$ is consistent with a current-induced inductive phase drop of $2\pi$ across the series inductance of the JVS.
		\textbf{(c, d)} Gain versus input phase and $\fdc$ at the degenerate point ($\fsignal = \fdc/2$). At lower RF power (c), the system exhibits a stable, nearly linear phase evolution. At higher RF power, comparing panel (d) to (b) shows that one period of gain modulation corresponds to a $2\pi$ shift of the squeezing angle. 
	}
\end{figure}

Naively, the ICTA response might be expected to remain invariant under changes to the externally applied bias \(h\fdc/2e\) once within the JVS locking range, as the voltage across the junction is rigidly fixed to the Shapiro step \(h\fJVS/2e\). However, as shown in Fig.~\ref{fig:fig3}, the squeezing angle exhibits an approximately linear evolution with \(\fdc\). This behavior originates from the phase drop generated by the dc current across the series inductance $L$ of the JVS.

When the JVS is locked, the voltage across it is fixed by the Josephson relation. If the externally applied bias differs from the locked voltage, the JVS draws a dc current $
I_\text{JVS}
=
\frac{h\left(\fdc-\fJVS\right)}{2eR}$,
where \(R=\SI{5}{\ohm}\) is the external bias source impedance. This current produces a phase drop 
\begin{equation*}
	\phi
	=
	\frac{2e}{\hbar} L I_\text{JVS}
	=
	2\pi \frac{L}{R}\left(\fdc-\fJVS\right)
\end{equation*}
across any series inductance \(L\) of the JVS, by which the effective phase presented to the ICTA is shifted. Sweeping the external bias, therefore, modulates the phase \(\phiJ\) at the ICTA interface, causing the squeezing angle to rotate even though the JVS voltage remains strictly locked.
Using the designed series inductance $L=\SI{10}{\nano\henry}$ of the low-pass filter, we obtain a slope of $\Delta \fdc=\SI{500}{\mega\hertz}$ per full phase turn, closely matching the experimental data. 

In Fig.~\ref{fig:fig4} we explore how increasing JVS RF power affects the JVS. At intermediate RF power ($\PJVS = \SI{-55}{\dBm}$) in panels (a) and (c), the phase evolves close to linearly, but the gain is significantly reduced  near the edge of the locking range. Increasing the JVS RF power further to $\PJVS = \SI{-43}{\dBm}$ broadens the Shapiro step, and the gain becomes strongly modulated with a period of $\SI{500}{\mega\hertz}$, as shown in Fig.~\ref{fig:fig4}(b,d). The modulation period corresponds to a $2\pi$ shift of the maximum and minimum gain with $\fdc$. We also note that the gain modulation is deeper at higher RF power (panels (b) and (d)). We attribute this effect to residual RF leakage from the JVS into the ICTA at high JVS RF power. Depending on the relative phase of the RF leakage with respect to the order-parameter phase difference across the junction, the RF tone enhances or reduces the effective Josephson energy available to the ICTA, thereby modulating the resulting gain. This interpretation is corroborated by semi-classical numerical simulations which reproduce the observed gain modulations (see \SIref{sec:Simulation}).

This strong interference at high JVS RF power confirms that the JVS phase is transferred to the ICTA via two channels, i.e.\ the superconducting order parameter and RF leakage. At lower JVS RF power, and therefore lower leaked RF power, we thus conclude that the phase reference is dominantly transferred by the superconducting order parameter. 


\begin{figure}
	\raggedright
	\includegraphics[width=1\columnwidth]{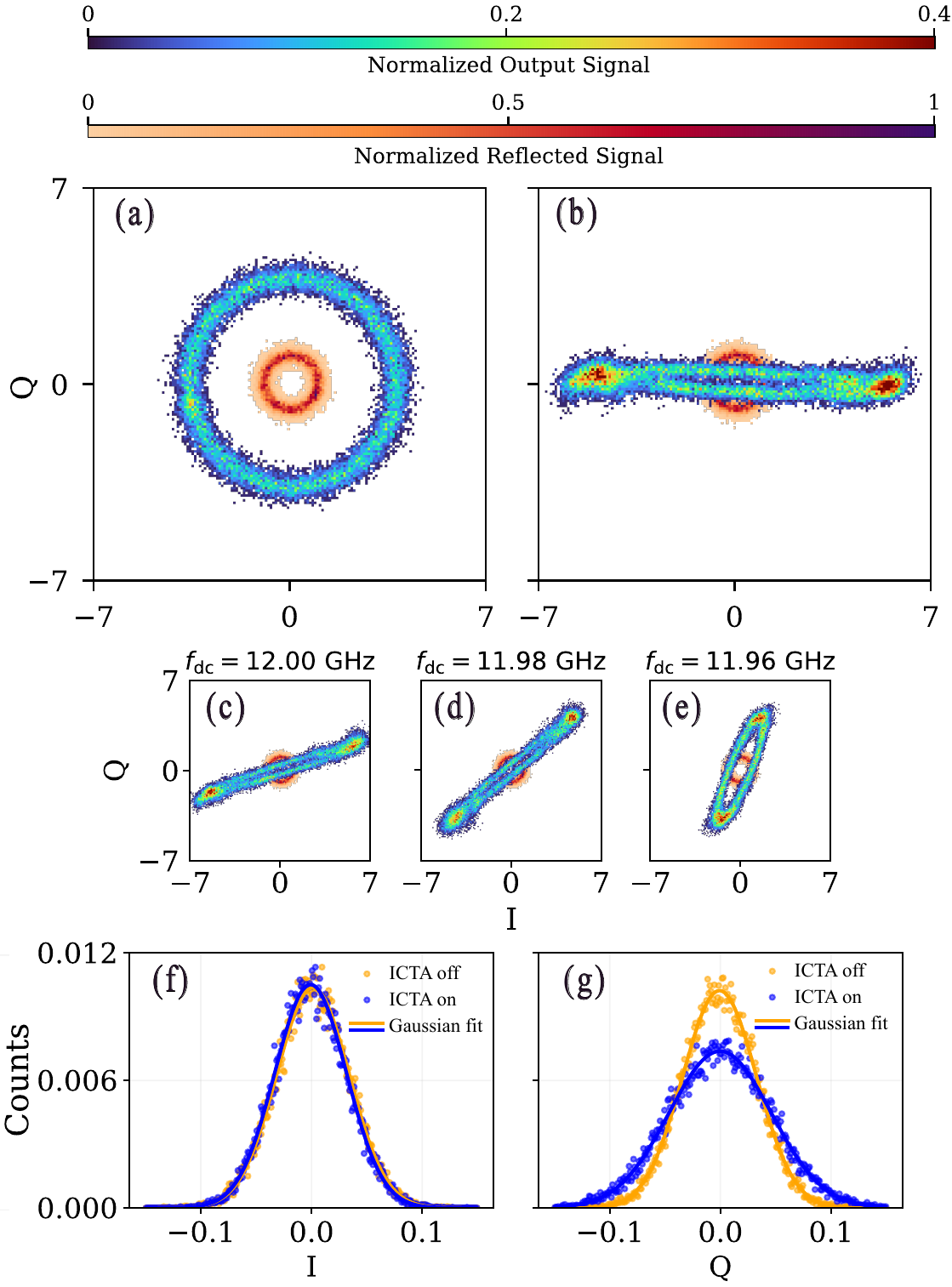}
	\caption{\label{fig:fig5}\textbf{System characterization in phase space.} In all panels, the input signal power is fixed at \(-120\,\mathrm{dBm}\), while its phase is swept. The reflected input state (red) is captured with both the ICTA and JVS deactivated. The JVS is driven by a $\SI{-65}{\dBm}$ RF tone. In panels (a–e), bins representing less than \(1\%\) of the peak histogram value are masked.
	\textbf{(a)} Phase-preserving amplification of a coherent state with the JVS deactivated. The amplified output (blue, ICTA active) exhibits a circular symmetric distribution, indicating uniform gain across both quadratures.
	\textbf{(b)} Phase-sensitive response with the JVS activated. The output distribution becomes elliptical, confirming quadrature-dependent amplification and de-amplification.
	\textbf{(c--e)} Phase-space responses at three different bias points, \(\fdc=12\), \(11.98\), and \(\SI{11.96}{\giga\hertz}\), demonstrating Josephson frequency control of the squeezing angle. Near the locking-range boundary (e), the squeezing is reduced due to increased phase noise from weakened voltage locking. 
	\textbf{(f,g)} Quadrature histograms for JVS-ICTA off (orange) and JVS-ICTA on (blue) without input signal. Data include the added noise of the HEMT in the amplification chain; solid curves represent Gaussian fits. In the squeezed quadrature, the variance is reduced by $\SI{2.8}{\percent}$ when the ICTA is turned on. In the orthogonal quadrature, the variance increases by a factor $1.91$.}
\end{figure}


To evaluate the system-level performance of the phase-stabilized ICTA, we characterize the phase-space distribution at the output of the full measurement chain including subsequent HEMT amplifiers and full down-conversion and digitization chain. In Fig.~\ref{fig:fig5} we show phase-space histograms where each count corresponds to a numerically down-converted phase-space amplitude filtered with a \(2^{16}\)-sample (\SI{32.8}{\micro\second} wide) flat-top window (see \SIref{sec:Phase-space Measurement}). 
We sweep the input signal phase at fixed power of \SI{-120}{\dBm} and acquire 256 points per phase. Measurements were taken both with the ICTA in the active state ($\fdc = \SI{12.02}{\giga\hertz}$) and the inactive state ($\fdc = \SI{1.1}{\giga\hertz}$), where it serves as a near-ideal reflector. 

Fig.~\ref{fig:fig5}(a) shows the phase-preserving response (JVS off) with the reflected input state (ICTA off) in red and the amplified output (ICTA on) in blue whcih both exhibit perfectly phase-symmetric distributions. The ratio of mean amplitudes for ICTA on and off gives a gain of \(\SI{9}{\dB}\). In contrast, activating the JVS ($\PJVS = \SI{-65}{\dBm}$) switches the ICTA into the phase-sensitive regime [Fig.~\ref{fig:fig5}(b)]. The resulting elliptical distribution confirms preferential amplification along one quadrature and deamplification along the orthogonal quadarature, corresponding to a phase-sensitive gain of \(\SI{17}{\dB}\) (see \SIref{sec:Phase-space Measurement} for additional gain settings).

As shown in Figs. \ref{fig:fig5}(c--e), the squeezing angle is deterministically rotated by shifting the applied voltage bias within the JVS locking range \(\fdc=12\), \(11.98\), and \(\SI{11.96}{\giga\hertz}\). This precise control confirms the successful transfer of the JVS phase reference to the ICTA via the superconducting order parameter. Near the edge of the locking range (\(\fdc=\SI{11.96}{\giga\hertz}\) in Fig.~\ref{fig:fig5}(e)), we observe a transition toward phase-preserving behavior and a reduction in squeezing, which we attribute to increased phase noise as the JVS voltage-locking weakens.

In order to verify that the JVS-ICTA indeed squeezes vacuum we compare noise emission with the JVS-ICTA input close to vacuum.
We first establish a reference with both the ICTA and JVS inactive, where the output of the ICTA is also close to vacuum. We then turn on the JVS-ICTA at \(\fJVS=\SI{12}{\giga\hertz}\) with power \(\SI{-65}{\dBm}\) to bias the ICTA at  \(\fdc=\SI{12}{\giga\hertz}\) and  $I_C=\SI{65}{\nano\ampere}$. For each measurement we acquire 25,600 histogram points with the same \SI{32.8}{\micro\second} wide flat-top window as before.
Figs.~\ref{fig:fig5}(f,g) present histograms for the squeezed and anti-squeezed quadratures. With respect to the reference, the Gaussian standard deviations with JVS-ICTA active of the sqeezed/antisqueezed quadrature change  by a factor $0.986$ and $1.382$, respectively. This corresponds to a small but significant variance reduction of \SI[separate-uncertainty = true]{2.8(6)}{\percent} in the squeezed quadrature, and the variance increases by a factor $1.91$ in the orthogonal quadrature. The HEMT noise, which is the dominant noise contribution in the measurement chain, limits the observed squeezing. Even for perfect sqeezing we would expect a reduction of $\frac{0.5}{N} \approx\SI{3.33}{\percent}$ with $N \approx 15$ the noise photon number of the HEMT amplifier refereed to the calibration point at the 6 port switch (see \SIref{sec:Wiring Setup}).

In summary, we have demonstrated that the inherent lack of phase reference of the voltage bias in Josephson photonics can be overcome by integrating a Josephson voltage standard (JVS) to establish a robust phase reference. Using the inelastic Cooper-pair tunneling amplifier (ICTA) as an example, we have shown that this phase reference is transferred via the superconducting order parameter and that it remains remarkably stable across several millimeters of on-chip circuitry, with the squeezing angle being precisely tunable via the bias-induced dc current.
This architecture thereby brings together the clean and Kerr-free nonlinearity of Josephson photonics with the phase-coherent operation required for many quantum information applications such as phase sensitive-amplification or stabilization of bosonic codes.

\bibliographystyle{apsrev4-2}
\bibliography{references}

	\clearpage
	\onecolumngrid
	
	\setcounter{page}{1}
	\setcounter{equation}{0}
	\setcounter{figure}{0}
	\setcounter{table}{0}
	
	\renewcommand{\theequation}{S\arabic{equation}}
	\renewcommand{\thefigure}{S\arabic{figure}}
	\renewcommand{\thetable}{S\arabic{table}}
	
	\renewcommand{\theHfigure}{S\arabic{figure}}
	\renewcommand{\theHequation}{S\arabic{equation}}
	\renewcommand{\theHtable}{S\arabic{table}}

\begin{center}
	\textbf{\large Supplementary Material}
\end{center}

\setcounter{page}{1}
\setcounter{equation}{0}
\setcounter{figure}{0}
\setcounter{table}{0}
\setcounter{section}{0}
\setcounter{subsection}{0}

\renewcommand{\theequation}{S\arabic{equation}}
\renewcommand{\thefigure}{S\arabic{figure}}
\renewcommand{\thetable}{S\arabic{table}}
\renewcommand{\thesection}{S\arabic{section}}

\renewcommand{\theHfigure}{S\arabic{figure}}
\renewcommand{\theHequation}{S\arabic{equation}}
\renewcommand{\theHtable}{S\arabic{table}}
\renewcommand{\theHsection}{S\arabic{section}}

\renewcommand\bibnumfmt[1]{[S#1]}
\renewcommand\citenumfont[1]{S#1}

\section{\label{sec:Fabrication} Fabrication }

Devices feature two niobium routing layers separated by SiN dielectric layer and aluminum Josephson junctions. They are fabricated on 3-inch sapphire wafers using \SI{100}{\kilo\volt} e-beam lithography in all steps. 

First a \(100~\mathrm{nm}\)-thick niobium ground plane is deposited by dc magnetron sputtering and patterned using \SI{300}{\nano\meter}-thick CSAR 62 AR-P~6200.13 resist and dry etching with \(\mathrm{SF_6 + Ar}\) (\(1:4\) flow ratio) ICP plasma.

Then \(200~\mathrm{nm}\)-thick silicon-nitride dielectric is deposited by plasma-enhanced chemical vapor deposition (PECVD) with a \(\mathrm{SiH_4:N_2}\) flow ratio of \(1:7\) and patterned using ma-N 2405 resist with  conductive Electra 92 AR-PC~5092 coating to reduce charging during exposure. Vias were opened using the same \(\mathrm{SF_6 + Ar}\) ICP etch.

After light ion milling, a second \(100~\mathrm{nm}\)-thick niobium film is deposited to form crossovers and parallel-plate capacitors. This layer was patterned using CSAR 62 AR-P~6200.13 and etched by a selective \(\mathrm{Ar + Cl_2}\) \((1{:}5)\) ICP plasma.

The final Josephson junction layers are fabricated using Manhattan-style shadow-angle evaporation of aluminium and lift-off with a trilayer resist stack consisting of MMA EL13 copolymer and a PMMA layer, with an additional Electra 92 coating for electron-beam exposure. Prior to aluminium deposition, the substrate is ion-milled, and the junction is formed by sequential deposition of \SI{40}{\nano\meter} of Al, dynamic oxidation at \(32~\mathrm{mT}\) with \(15\,\mathrm{sccm}\) of \(\mathrm{O}_2\) for \(15\,\mathrm{min}\), and \SI{70}{\nano\meter} of Al to form the second junction lead. After lift-off, the wafer is coated with protective dicing resist and diced into \(5\times5\,\mathrm{mm^2}\) chips.

\section{\label{sec:Device Design} Device Design}

\subsection{\label{sec:Josephson Voltage Standard Design} Josephson Voltage Standard Design (JVS)}
The Josephson voltage standard is described by the resistively and capacitively shunted junction (RCSJ) model. In the small-phase limit, the junction behaves as a linear resonator with plasma frequency $f_p=\sqrt{2eI_c/\hbar C}$, Josephson inductance $L_J \approx \hbar/(2eI_c)$, and quality factor $Q=R\sqrt{C/L_J}$, where in our case $R=\SI{50}{\ohm}$ is the transmission line impedance of the RF drive to the JVS. Our design aims for underdamped operation to minimize on-chip dissipation. 

For well-developed Shapiro spikes avoiding chaotic dyanmics, the requirements are typically \(Q>1/2\) and \(\fJVS>f_p\), where \(\fJVS\) is the drive frequency. In normalized form,  these conditions are written as \(F_R\gg 1\) and \(F_R\gg 1/\sigma\), where \(F_R=\fJVS/f_p\) and
\(\sigma = R^{-1}\left({\hbar}/{I_c C 2e}\right)^{1/2}\)\cite{kautz1996noise}. 

Josephson photonic circuits typically require dc voltages of a few tens of microvolts and currents of up to several hundred nanoamperes. The JVS was designed to provide 5 to \SI{50}{\micro\volt} and output currents up to approximately \SI{1}{\micro\ampere}. Therefore, the final design of the JVS employs a junction with  \(I_{c,JVS} \approx \SI{3}{\micro\ampere}\) and a shunt capacitance of approximately \SI{40}{\pico\farad}, yielding a plasma frequency of \(f_p \approx \SI{1.5}{\giga\hertz}\). The circuit includes an on-chip bias tee for simultaneous dc biasing and RF driving, as well as a large shunt capacitor \(C_P=\SI{300}{\pico\farad}\) as the final stage of the \SI{5}{\ohm} voltage biasing circuit \cite{martel2025influence}.

\begin{figure}[H]
	\centering
	\includegraphics[width=0.55\textwidth]{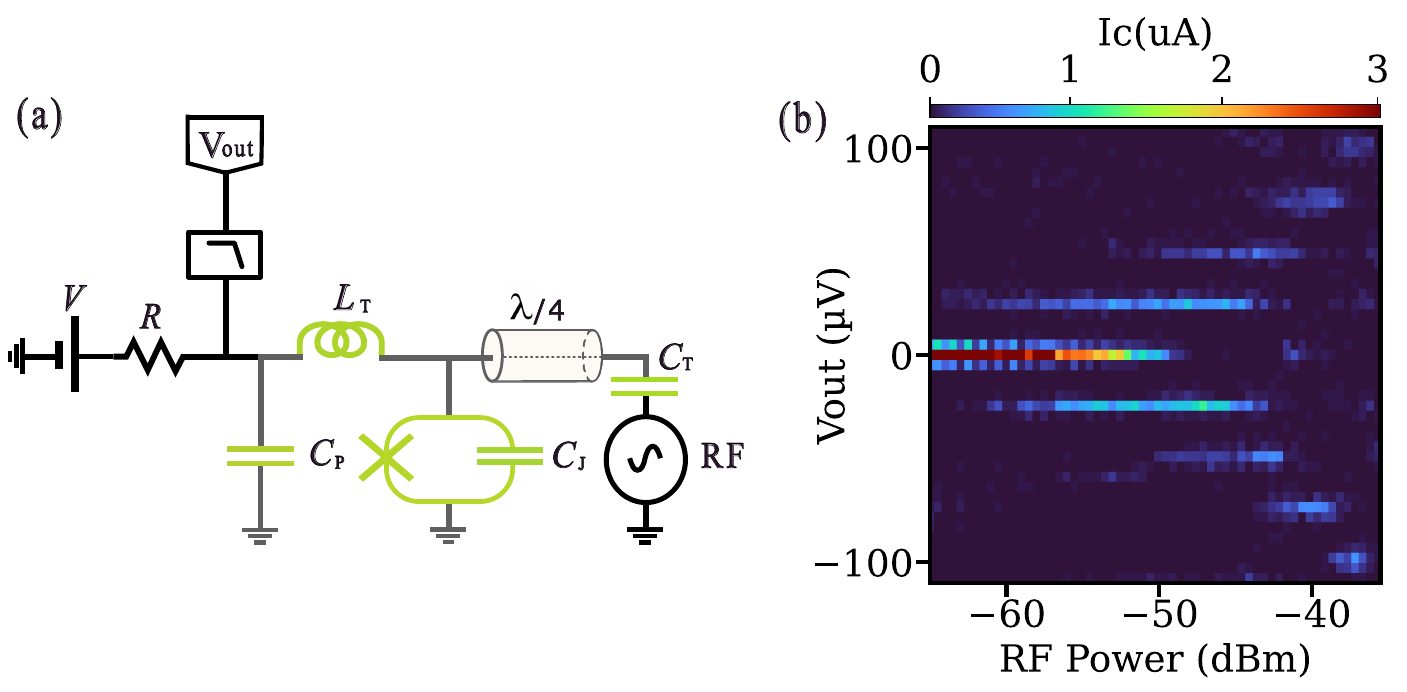}
	\caption{\label{fig:figS1}
		\textbf{Standalone JVS design and Shapiro-spike measurements.}
		\textbf{(a)} Schematic representation of the calibration JVS device. Unlike the JVS integrated with the ICTA in the main experiment, this device includes a \(\lambda/4\) impedance transformer, which transforms the \(50~\Omega\) microwave environment to an effective impedance of approximately \(30~\Omega\) at the Josephson junction. The Josephson junction is shunted by the capacitance \(C_J\), with an on-chip bias-tee formed by \(L_T\) and \(C_T\), and the dc output is monitored at \(V_{\mathrm{out}}\).
		\textbf{(b)} A dc characterization of the JVS as a function of incident RF power. A low-frequency triangular waveform, described in \SIref{sec:DC Measurement Setup}, is used as the dc bias to reveal the quantized voltage plateaus corresponding to Shapiro steps induced by the RF drive. The color scale indicates the maximum dc current sustained by each Shapiro step.}
\end{figure}

For the standalone JVS device (Fig.~\ref{fig:figS1}), a \(\lambda/4\) impedance transformer at \SI{2}{\giga\hertz} with characteristic impedance \SI{39}{\ohm} transforms the \SI{50}{\ohm} microwave environment to an effective impedance of approximately $R=\SI{30}{\ohm}$ at the JVS junction. This low-impedance environment improves the stability of the Shapiro steps and enables stablization of voltage plateaus down to approximately \SI{4}{\micro\volt}, with \(\sigma\approx0.05\). In the integrated ICTA-JVS device used for the measurements in the main manuscript, the \(\lambda/4\) transformer is omitted, and the junction instead sees an effective impedance  $R=\SI{50}{\ohm}$. This configuration is sufficient because the JVS is operated around \SI{12}{\giga\hertz}, well above its plasma frequency.

Fig.~\ref{fig:figS1}(a) illustrates the standalone JVS layout, with component values detailed in Table~\ref{tab:tab1}. DC characterization is performed using the setup described in \SIref{sec:DC Measurement Setup} by measuring (\(V_{\mathrm{out}}\)) when applying a triangular dc voltage $V$, together with an RF tone with frequency \(\fJVS=\SI{12}{\giga\hertz}\). This measurement provides a calibration of the RF power required to stabilize each ($n\times\SI{24}{\micro\volt}$) spike, and the maximum dc current it supports (see Fig.~\ref{fig:figS1}(b)). The devices are simulated by \textsc{WRspice} and \textsc{ADS} which showed excellent agreement with the experimental response.

\begin{table}[H]
	\centering
	\renewcommand\arraystretch{1.3}
	\caption{\label{tab:tab1}JVS component values corresponding to Fig.~\ref{fig:figS1}.}
	\begin{tabular}{|*{7}{c|}}
		\hline
		$C_P$  & $R$  & $\lambda/4$ impedance transformer & JJ max.\ $I_c$  &  $C_J$  &   $C_T$   &   $L_T$   
		\\\hline
		\SI{300}{\pico\farad}  & \SI{5}{\ohm}  & \SI{39}{\ohm}, \SI{2}{\giga\hertz} & \SI{3}{\micro\ampere} & \SI{30}{\pico\farad} & \SI{40}{\pico\farad} & \SI{10}{\nano\henry}
		\\\hline
	\end{tabular}
\end{table}

\subsection{\label{sec:Filter Design} Filter Design}
\begin{figure}[H]
	\centering
	\includegraphics[width = 0.65\textwidth]{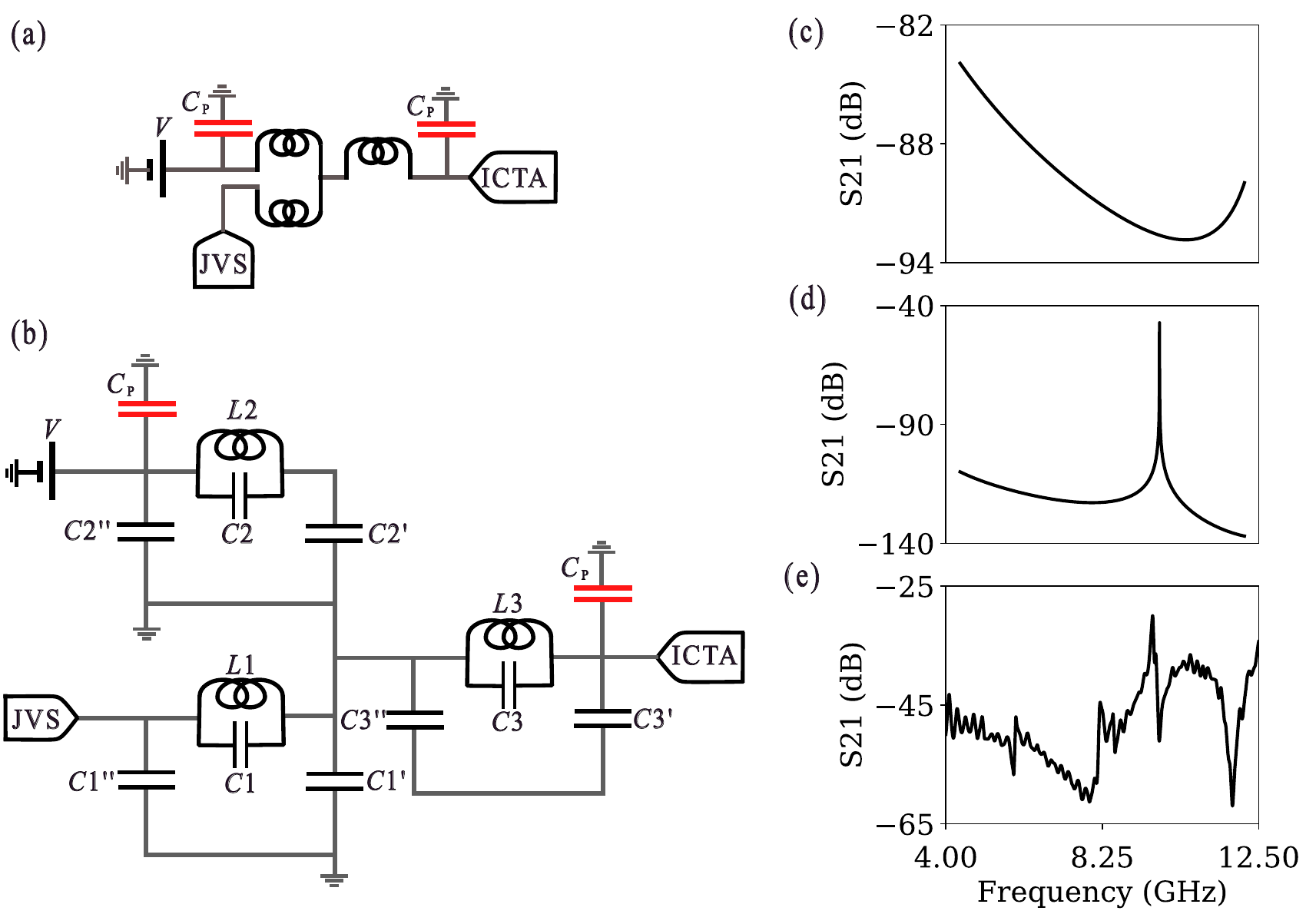}
	\caption{\label{fig:figS3}
		\textbf{Filter design and performance}
		\textbf{(a)} Schematic of the superconducting filter connecting the JVS and ICTA.
		\textbf{(b)} Expanded circuit model, in which each spiral inductor is described by a \(\pi\)-network including parasitic elements.
		\textbf{(c)} Simulated transmission of the filter alone, showing strong isolation over the \SI{4}{\giga\hertz}--\SI{12.5}{\giga\hertz} range.
		\textbf{(d)} Simulated transmission from JVS RF port to ICTA input/outout port through the filter, showing further improved isolation together with a parasitic resonance near \SI{10}{\giga\hertz}. 
		\textbf{(e)} Measured transmission of the filter, showing an average isolation of
		\(\sim \SI{50}{\decibel}\), reaching \(\sim \SI{65}{\decibel}\) at \SI{12}{\giga\hertz}.}
	
\end{figure}

The schematic of the superconducting low-pass filter connecting the JVS to the ICTA is shown in Fig.~\ref{fig:figS3}(a). Its topology increases the RF impedance of the shared bias line while preserving a continuous dc path. Large shunt capacitors \(C_P\)  at the input dc port and the ICTA-side port provide supplemental low-pass filtering and suppress RF leakage from the JVS to the ICTA. Collectively, the series inductance and shunt capacitance route RF leakage to ground and reduce parasitic coupling through the bias line without perturbing the dc bias. This design ensures that filter self-resonances and leakage remain outside the primary ICTA operating band. Crucially, the superconducting path enables transfer of the phase reference from the JVS to the ICTA via the superconducting order parameter.

The layout was optimized using schematic simulations in \textsc{ADS} with each inductance modeled by a \(\pi\)-network (see Fig.~\ref{fig:figS3}(b)) with parasitic shunt capacitances \((C_1,C_2,C_3)\) and capacitances to ground \((C_1',C_1'',C_2',C_2'',C_3',C_3'')\) extracted from HFSS simulations. The simulations predict isolation of approximately \(80\,\mathrm{dB}\) over the \SI{4}{\giga\hertz}--\SI{12.5}{\giga\hertz} band (see Fig.~\ref{fig:figS3}(c)). A full-device simulation, containing JVS and ICTA, indicated further improved isolation up to \(100\,\mathrm{dB}\), with a parasitic resonance near \SI{10}{\giga\hertz}, outside the operating band of our experiments (see Fig.~\ref{fig:figS3}(d)) due the transmission line between the filter and ICTA.

Experimentally, the filter provides an average isolation of \(\sim 50\,\mathrm{dB}\), reaching \(\sim 65\,\mathrm{dB}\) at \SI{12}{\giga\hertz}, where the JVS stabilizes the ICTA bias (see Fig.~\ref{fig:figS3}(e)). To characterize the isolation between the JVS and ICTA, we first calibrated the output line using the Y-factor method (see Sec.~\ref{sec:Noise and SNA Measurement}). The ICTA was then operated in unity-reflection mode, where it acts as an approximately ideal reflector, and the resulting measurement was used as a calibration for the output line. Next, a signal was applied through the JVS RF port and coupled to the ICTA, while the ICTA remained in unity-reflection mode, and the signal was measured at the output port. Thus, this measurement characterizes the transmission of a signal from the JVS RF port through the ICTA and into the output line, rather than through the conventional ICTA input line. 
In this measurement, the JVS RF line has not been directly calibrated, but has the same geometry and materials as the ICTA input line. We therefore suppose that the JVS RF line and the ICTA input line have the same attenuation except for the discrete attenuators. We compensate the discrete attenuators on the ICTA input line inside the frige, by placing the same attenuation value on JVS RF line at room temperature. We expect the imbalance between the lines to be less than \SI{5}{\decibel}.

The discrepancy between the measured and simulated isolation (Fig.~\ref{fig:figS3}(c-e)) is likely due to chip- and package-level electromagnetic coupling that is not captured by the schematic model. Filter component parameters are detailed in Table~\ref{tab:tab3}.

\begin{table}[H]
	\centering
	\renewcommand\arraystretch{1.3}
	\caption{Filter components values (Fig.~\ref{fig:figS3})}
	\definecolor{lightgray}{gray}{1}
	\label{tab:tab3}
	\begin{tabular}{|*{13}{c|}}
		\hline
		$C_P$  & $L1'$ & $C1$ & $C1'$  & $C1"$   & $L2$ & $C2$ &  $C2'$  & $C2"$  & $L3$ & $C3$ &   $C3'$  & $C3"$\\
		\hline
		\SI{300}{\pico\farad}  &  \SI{10}{\nano\henry} & \SI{7}{\femto\farad} &	\SI{10}{\femto\farad} &\SI{14}{\femto\farad} &  \SI{5}{\nano\henry}  &  \SI{5.5}{\femto\farad} &	\SI{8}{\femto\farad} & \SI{12}{\femto\farad} &  \SI{10}{\nano\henry}  &  \SI{7}{\femto\farad} &	\SI{10}{\femto\farad} & \SI{14}{\femto\farad} \\
		\hline
	\end{tabular}
\end{table}

\subsection{\label{sec:Inelastic Cooper-pair Tunneling Amplifier Design} Inelastic Cooper-pair Tunneling Amplifier Design (ICTA)}

\begin{figure}[H]
	\centering
	\includegraphics[width = 0.6\textwidth]{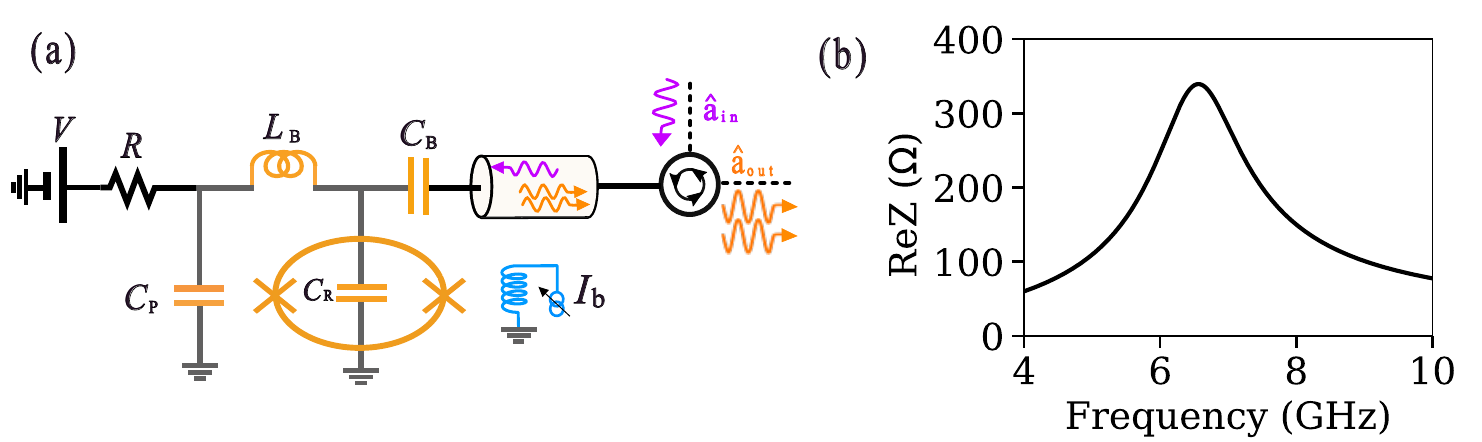}
	\caption{\label{fig:figS2}\textbf{ICTA design and Impedance seen by SQUID}
		\textbf{(a)} Schematic of the ICTA in orange, consisting of a SQUID embedded in a single wide-band resonant mode formed by \(L_B\), \(C_B\), and the shunt capacitance \(C_R\). The mode is designed near \SI{6}{\giga\hertz}, with a peak real impedance of about \SI{300}{\ohm} and a bandwidth of approximately \SI{1.2}{\giga\hertz}. The blue inductor indicates the flux-bias for the SQUID. 
		\textbf{(b)} Real part of impedance seen by the SQUID, calculated using \textsc{ADS} and our custom ICTA simulator impedance and gain model. The plotted impedance is extracted from the ICTA simulator model.}
\end{figure}

The ICTA\cite{nehra2026broadband} is built around a single wide-band resonant mode, formed by the bias-tee inductance \(L_B\) and capacitance \(C_B\), and an additional shunt capacitance \(C_R\) placed in parallel with the SQUID (see Fig.~\ref{fig:figS2}). This resonator is designed near \SI{6}{\giga\hertz}, with a maximum real impedance of \SI{300}{\ohm} and a bandwidth of about \SI{1.2}{\giga\hertz}. This design hosts both signal and idler modes in a single resonator, enabling both non-degenerate and degenerate operation. The circuit also includes a shunt capacitor \(C_P\) to ground, as in the JVS, which acts as the final on-chip low-pass filtering stage for the dc bias line.

Fig.~\ref{fig:figS2}(a) shows the ICTA schematics, highlighted in orange. The gain is adjustable by using a SQUID as adjustable Josephson element (flux biasing idicated in blue). Fig.~\ref{fig:figS2}(b) shows the impedance of the linear part of the circuit, as seen from the SQUID, obtained from both \textsc{ADS} simulations and our custom ICTA impedance and gain simulator~\cite{nehra2026broadband}. The impedance calculated by our custom simulator agrees well with the \textsc{ADS} results. This impedance is then used by the custom ICTA gain simulator to calculate the gain, which also shows good agreement with the experimental measurements.

\begin{table}[H]
	\centering
	\renewcommand\arraystretch{1.3}
	
	\caption{ICTA components values  (Fig.~\ref{fig:figS2})}
	\label{tab:tab2}
	\begin{tabular}{|*{6}{c|}}
		\hline	
		$C_P$  & $R$  &  Max. $I_c$ (per JJ)  &  $C_R$  &   $C_B$   &   $L_B$ \\
		\hline
		\SI{300}{\pico\farad}  & \SI{5}{\ohm}  &  \SI{0.3}{\micro\ampere} &	\SI{220}{\femto\farad} & \SI{210}{\femto\farad} &   \SI{1.5}{\nano\henry}\\
		\hline
	\end{tabular}
\end{table}

\section{\label{sec:Measurements Configuration} Measurements Configurations}

\subsection{\label{sec:Wiring Setup} Measurements wiring}
\begin{figure}[t]
	\centering
	\includegraphics[width = 0.85\textwidth]{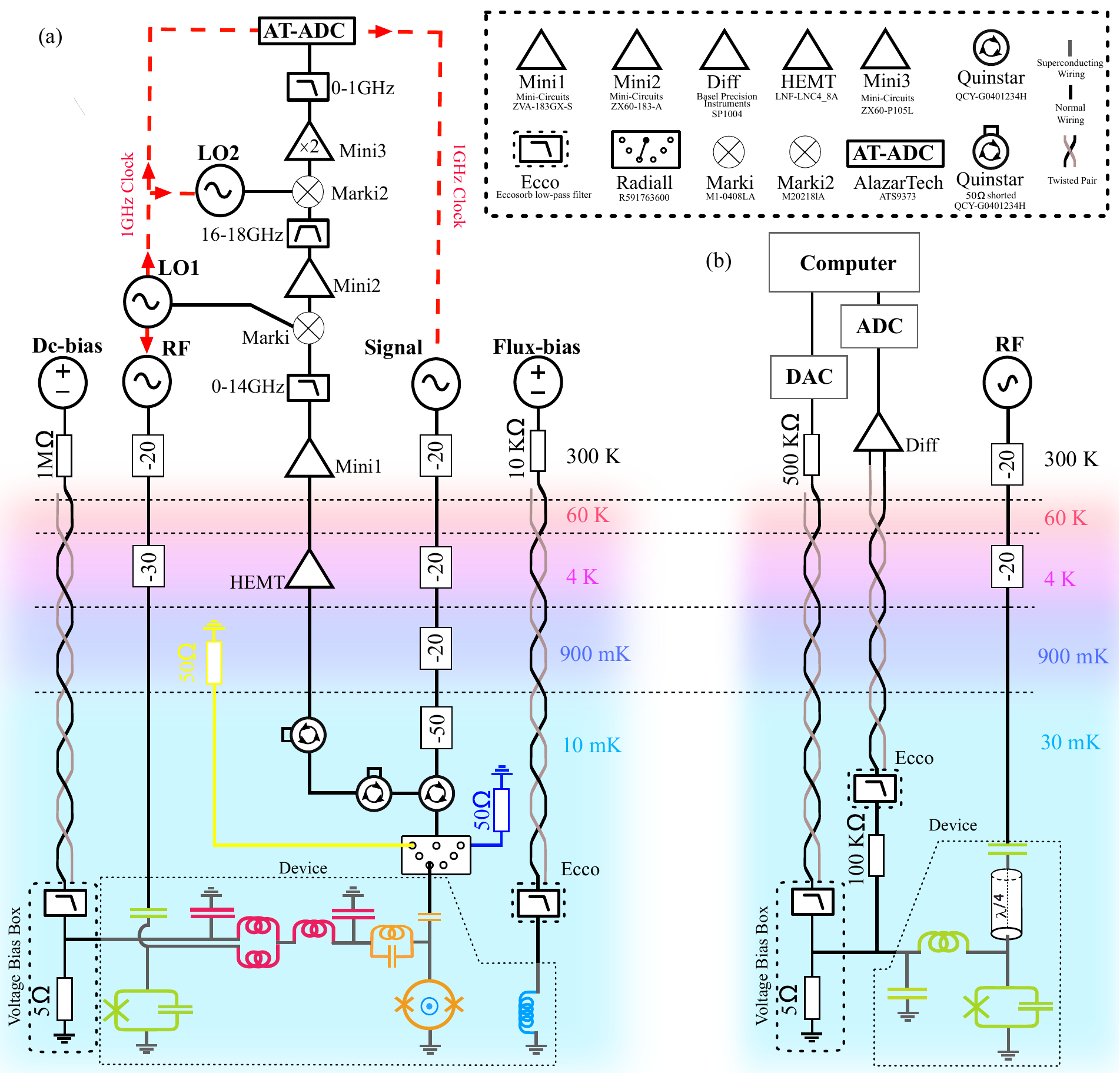}
	\caption{\label{fig:figS4}
		\textbf{Wiring configurations used for the measurements in the two dilution refrigerators.}
		In both setups, the voltage-bias box, device, and Eccosorb filter are connected to the refrigerator ground. \textbf{(a)} Cryogenic microwave measurement setup for characterizing of the ICTA. From left to right, the first line supplies the dc bias through a resistive divider formed by a \(500~\mathrm{k}\Omega\) resistor and the \(5~\Omega\) bias box. The second line provides the RF drive tone for the JVS with a total attenuation of about \(40\,\mathrm{dB}\) inside the fridge. The third line is the output chain, which includes a triple junction isolator, a HEMT amplifier at \SI{4}{\kelvin} and a room-temperature double-heterodyne receiver. The fourth line is the signal line which is attenuated at different temperature stages and routed to the device through a circulator and a six-port switch. For Y-factor noise calibration, two \SI{50}{\ohm} terminations are anchored at the mixing chamber (dark blue) and still plate (yellow). The final line provides the dc flux bias through twisted-pair wiring and an Eccosorb low-pass filter.
		\textbf{(b)} Cryogenic dc measurement setup for the standalone JVS. From left to right, the first line is used to apply a triangular low-impedance bias signal through a resistive divider. The second line is used to observe the voltage \(V_{\mathrm{out}}\) at the JVS. An Eccosorb low-pass filter and a \(100~\mathrm{k}\Omega\) series resistor isolate the measurement line from RF leakage and reduce back-action on the device. The signal is read out through differential twisted-pair wiring, amplified by a low-noise differential amplifier, and digitized with a 16-bit ADC.}
	
\end{figure}

Fig.~\ref{fig:figS4}(a) shows the measurement wiring used in the main part of this work and Fig.~\ref{fig:figS4}(b) the one used for the standalone JVS measurement presented in Fig.~\ref{fig:figS1}. The setups comprise the following components:

\textbf{Flux-bias line:} 
The flux-bias circuit (Fig.~\ref{fig:figS4}(a), right) provides an external magnetic flux to the SQUID loop to tune the Josephson energy and control the ICTA gain. The bias is supplied by a high-precision source (Bilt BN103) through a \SI{10}{\kilo\ohm} series resistor at room temperature. The dc current is routed to the mixing chamber via twisted-pair wiring and passes through an Eccosorb CR-124 low-pass filter \cite{paquette2022absorptive} at the \SI{10}{\milli\kelvin} stage.

\textbf{Signal line / readout chain:}
The input signal line, adjacent to the flux-bias line in Fig.~\ref{fig:figS4}(a), delivers the microwave input signal to the ICTA. The signal is generated by a Rohde \& Schwarz SGS100A source and undergoes \SI{20}{\decibel} of attenuation at room temperature, followed by tiered attenuation at the \SI{4}{\kelvin} stage, the \SI{900}{\milli\kelvin} stage, and the base plate, ensuring the input line has minimal thermal photon population. The total attenuation inside the dilution refrigerator is approximately \SI{100}{\decibel}. Then, the signal passes through a cryogenic circulator before reaching a six-port switch, which selectively routes the signal to the sample, a short circuit, or two matched loads anchored to the mixing chamber and still stage of the dilution refrigerator. 
The signal reflected by the device is routed to a cryogenic HEMT amplifier (LNF LNC0.3\_14A) by the circulator. We use a triple-junction circulator to suppress amplifier back-action. 

Outside the refrigerator, the signal is then further amplified by a room-temperature amplifier (Mini-Circuits ZVA-183GX-S+) and downconverted via a double-heterodyne conversion setup. In the first mixing stage (Marki MM1-0312S), the reflected signal mixed with a tunable local oscillator (LO1) between $22$ and $\SI{30}{\giga\hertz}$ generated by a Rohde \& Schwarz SMF100A. The output of this first stage is bandpass filtered between $16$ and $\SI{18}{\giga\hertz}$. In the second mixing stage, the signal is down-converted to base band by a second mixer (Marki M202181A) driven by a fixed \SI{16}{\giga\hertz} local oscillator (LO2) generated from the \SI{1}{\giga\hertz} reference oscillator of LO1 using a comb generator, band pass filters and amplifiers. Following final amplification and anti-aliasing filtering, the signal is digitized with a 12-bit AlazarTech ATS9373 analog-to-digital converter operating at \SI{2}{GS/s}, clocked by the second harmonic of the LO1 reference oscillator.

\textbf{RF line (RF tone for JVS):}
The RF tone used to drive the JVS is generated by a Rohde \& Schwarz SGS100A phase-locked to the \SI{1}{\giga\hertz} reference oscillator of LO1. The RF tone is attenuated by \(20\,\mathrm{dB}\) at room temperature and \(30\,\mathrm{dB}\) at the 4~K stage, and then applied directly to the device. 

\textbf{dc-bias line and bias box:} 
The dc-bias circuit, located on the far left of Fig.~\ref{fig:figS4}(a), provides a stable bias $\Vdc$ for the JVS. The bias is generated by a high-precision source (Bilt BN103) in series with a \SI{1}{\mega\ohm} resistor at room temperature, forming a voltage divider with a \SI{5}{\ohm} resistor at the mixing stage. The cold resistor is integrated with a low-pass filter with flat \SI{5}{\ohm} output impedance up to approximately \SI{1}{\giga\hertz}. The final stage of low-pass filtering is implemented on-chip by a large shunt capacitor \(C_P\) to ground. Collectively, this tiered architecture suppresses voltage noise on the bias line and shields the device from high-frequency parasitics originating from the dc line \cite{martel2025influence}.

\subsection{\label{sec:Noise and SNA Measurement} Noise and SNA Measurement}

To accurately determine the gain and added noise of the ICTA, we perform a multi-step calibration procedure. In a first step we calibrate gain and noise of the readout chain using the Y-factor method. We measure the power spectral density (PSD) emitted by two \SI{50}{\ohm} terminations anchored at different temperature stages. One termination is thermalized at the mixing chamber plate (blue resistor, Fig.~\ref{fig:figS4}(a)) and the other at the still plate (yellow resistor, Fig.~\ref{fig:figS4}(a)). Neglecting the attenuation of the short superconducting coax sections between the terminations and the switch, the reference point of this calibration is the 6 port switch.

The second calibration step determines the input-line attenuation. For this purpose, the cryogenic switch is configured open (all ports disconnected), and a SNA measurement is performed to obtain the total transmission from the RF input to the output of the chain. By combining this result with the previously calibrated output chain gain, the total input attenuation to the swtich is extracted, which is approximately \SI{90}{\decibel}.

Next, the switch is connected to the sample port, and the ICTA is biased at a point of unity gain (by adjusting the dc and flux bias). The square root of the ratio of this reflection to reflection at the switch gives the transmission coefficient of cables and components between the switch and the device. This measurement serves as the final reference for the ICTA gain characterization. The intrinsic gain of the ICTA is then measured via the SNA method and normalized against this device-level reference.

To characterize the added noise of the amplifier, we measure the PSD of the spontaneous emission of the device. The measured output PSD is first referred to the ICTA output using the calibrated output readout chain gain and then referred to the ICTA input using the calibrated device gain. The resulting input-referred noise is used to determine the added noise of the ICTA.

\subsection{\label{sec:Phase-space Measurement} Phase-space Measurement}

To visualize the IQ histograms of the emitted and reflected signals, we employed the double-heterodyne configuration shown in Fig.~\ref{fig:figS4}(a). With RF, LO1, LO2 and digitizer phase-locked, we choose the RF signal frequency such that 
the reflected signal is downconverted to \SI{0.5}{\giga\hertz} at the center of the first Nyquist zone. This frequency relation ensures consistent phase relations irrespective of when the acquisition is triggered with respect to the reference clock. We sweep the rf tone phase with respected to the reference clock phase and keep the LO1, LO2, digitizer clock phases fixed.

 Each datapoint in the IQ histogram corresponds to the complex amplitude extracted from a block of \(2^{16}\) samples, corresponding to an integration time of \SI{32.8}{\micro\second}, digitally demodulated and flattop windowed.  We sweep the rf tone phase over $2\pi$ in 70 steps and take 256 of such data points at each phase angle. 

To enhance the clarity of the overlaid IQ histograms in Fig.~\ref{fig:fig5}, a threshold mask was applied: bins with counts below \(1\%\) of the peak histogram value were suppressed to remove weak tails and background statistical noise. Using transparent colormaps with a white background allows the statistically dominant portions of the distributions to remain prominent, facilitating a direct comparison of their relative morphologies.

\subsection{\label{sec:DC Measurement Setup} DC Measurement Setup}

The dc measurement setup used for JVS calibration in Fig.~\ref{fig:figS1}(b) is shown in Fig.~\ref{fig:figS4}(b). A triangular dc voltage is generated by a digital-to-analog converter (DAC) and delivered to the device through the same divider and filtering circuit as for the main experiment. The output voltage $V_{\mathrm{out}}$ of the JVS, is monitored as a function of dc bias and RF tone. The $V_{\mathrm{out}}$ signal is routed through a \SI{100}{\kilo\ohm} series resistor and an Eccosorb low-pass filter \cite{paquette2022absorptive} to minimize back-action on the JVS.
$V_{\mathrm{out}}$ is routed via differential twisted-pair wiring and amplified at room temperature using a low-noise, low-drift differential amplifier (Basel Precision Instruments SP1004). The amplified signal is then digitized with a 16-bit analog-to-digital converter (ADC).

\section{\label{sec:Simulation and Theory} Auxiliary calculations, simulations, and measurements on the ICTA}
\subsection{\label{sec:Phase-sensitive and phase-preserving regimes of the single cavity ICTA}Phase-sensitive and phase-preserving regimes of the single cavity ICTA}

The single-cavity ICTA used in this work is described by the Hamiltonian

\begin{equation}
	H = \hbar \omega_a a^\dagger a - E_J \cos\left(\omega_{\mathrm{dc}} t + \phiJ + \phi_a \left(a^\dagger + a\right)\right),
\end{equation}

where $E_J$ is the Josephson energy, and $\omega_{\mathrm{dc}}$ is the Josephson angular frequency $(\omega_{\mathrm{dc}} = 2eV_{\rm dc}/\hbar)$, and $\phiJ$ is the phase associated with the voltage bias. The parameter $\phi_a = \sqrt{\pi \frac{4e^2}{h}Z_a}$ is the zero-point fluctuation of the phase of mode $a$ at frequency $\omega_a$, where $Z_a$ is the characteristic impedance of the mode.

For small signals and small $\phi_a$, the Josephson term can be expanded to second order in $\phi_a)$. Under the rotating-wave approximation (RWA) at $\omega_J \approx 2\omega_a$, the Hamiltonian can then be approximated

\begin{equation}
H \approx \hbar \omega_a a^\dagger a + \hbar\xi^* e^{i\omega_{\mathrm{dc}} t} a^2 + \hbar\xi e^{-i\omega_{\mathrm{dc}} t} {a^\dagger}^2.
\end{equation}

Coupling to a transmission line with coupling rate $\gamma$ is described by the quantum Langevin equation 
\begin{equation}
\frac{\text{d}a(t)}{\text{d}t} = \frac{i}{\hbar} \left[H,a(t)\right] - \frac{\gamma}{2} a(t) + \sqrt{\gamma} \ain(t)
\end{equation}
with boundary condition 
\begin{equation}
\ain(t) + \aout(t) = \sqrt{\gamma} a(t).
\end{equation}
We Fourier transform the quantum Langevin equation and use the boundary condition to eliminate the cavity mode and get a scattering  relation
\begin{equation}
\dets \aout(\omega) + i\xi \aout^\dagger(\omega_{\mathrm{dc}}-\omega) = \dets^* \ain(\omega) - i \xi \ain^\dagger(\omega_{\mathrm{dc}}-\omega)
\end{equation}
for the spectral modes
\begin{equation}
a_\text{in/out}(\omega) = \frac{1}{\sqrt{2\pi}} \int a_\text{in/out}(t) e^{i\omega t} \text{d} t
\end{equation} 
with $\dets = \frac{\gamma}{2} - i (\omega - \omega_a)$. 
Its solution is
\begin{equation}
\aout(\omega) = u \ain(\omega) + v \ain^\dagger(\omega_{\mathrm{dc}}-\omega)
\label{eq:gain}
\end{equation}
with
\begin{eqnarray}
u &=& \frac{\dets^* \deti^* + |\xi|^2}{\dets \deti^* - |\xi|^2}\\
v &=& \frac{-i\xi\gamma}{\dets \deti^* - |\xi|^2}
\end{eqnarray}
where $\deti = \frac{\gamma}{2} - i (\omega_{\mathrm{dc}} - \omega - \omega_a)$.

In the non-degenerate case $2 \omega \neq \omega_{\mathrm{dc}}$, Eq.~\eqref{eq:gain} leads to phase preserving amplification with amplitude gain $u$. It leads to quantum-limited noise when all input modes are in the ground state and commute: 
\begin{equation}
 \langle\aout^\dagger(\omega) \aout(\omega)\rangle = |v|^2 = |u|^2-1
\end{equation}

We focus here on the degenerate case $\omega =  \omega_{\mathrm{dc}}/2$. To simplify notation we also suppose $\omega = \omega_a$. Under these conditions $u$ and $v$ simplify to 
\begin{eqnarray}
u &=& \frac{1+ |\Xi|^2}{1- |\Xi|^2} = \cosh r\\
v &=& \frac{-2i\Xi}{1- |\Xi|^2} = e^{-i(\phiJ+\frac{\pi}{2})} \sinh r  
\label{eq:vdegenerate}
\end{eqnarray}
with $\Xi = 2\xi/\gamma = 2\phi_a^2 \frac{\Ej}{\hbar\gamma}e^{-i\phiJ}$.
$\omega \neq \omega_a$ would simply add a phase offset in $\Xi$. 

The response to a coherent state with $\langle \ain \rangle = A e^{i\theta}$ is 
\begin{align}
	\langle\aout\rangle
	&=
	A\left(u e^{i\theta}+v e^{-i\theta}\right)
	\nonumber\\
	&= 
	A\left(\cosh r\,e^{i\theta}+e^{-i(\theta+\phiJ+\frac{\pi}{2})} \sinh r\right) 
	\nonumber\\
	&=g  \langle \ain \rangle
	\label{eq:aout_factor_supp}
\end{align}
with 
\begin{eqnarray}
g &=& \cosh r+e^{i\Delta} \sinh r\\
-\Delta &=& \phiJ+\frac{\pi}{2}+2\theta
\end{eqnarray}
The magnitude and phase of the gain depend on input phase $\theta$ and the phase reference $\phiJ$:
\begin{eqnarray}
	|g| &=& \sqrt{\left(\cosh r+\sinh r\cos\Delta\right)^2
	+\left(\sinh r\sin\Delta\right)^2} = \sqrt{\cosh 2r+\sinh 2r \cos\Delta}\\
	\label{eq:aout_magnitude_sq}
\arg g &=& \arg\!\left(\cosh r+\sinh r\,e^{i\Delta}\right)
	= \arctan\!\left(
	\frac{\sinh r\,\sin\Delta}{\cosh r+\sinh r\,\cos\Delta}
	\right)
	\label{eq:Phi_supp}\\
\end{eqnarray}

This explains the phase behavior observed experimentally in the degenerate regime (Fig.~\ref{fig:fig3}): unlike the non-degenerate (phase-preserving) case the degenerate gain phase depends on the signal phase with respect to the phase $\Delta$ and is $\pi$-periodic in $\theta$ and $2\pi$ periodic in $\phiJ$. The gain is maximal, $g= e^r > 1$, for $\Delta = 0$ and minimal, $g=e^{-r} < 1$, for $\Delta=\pi$ close to this minimum, the $\cosh$ and $\sinh$ terms interfere destructively, and the gain magnitude and phase become highly sensitive on the signal phase $\theta$ as can be seen in Fig.~\ref{fig:fig3}(b,c).

When the JVS is turned off, the ICTA phase $\phiJ$ diffuses due to voltage noise. In our experiments we typically see a distribution of $\omega_{\mathrm{dc}}$ larger than $2\pi \cdot \SI{5}{\mega\hertz}$, i.e.\ the phase correlation function decays of a timescale $\left(2\pi \cdot \SI{5}{\mega\hertz}\right){-1}\approx \SI{30}{\nano\second}$, 1000 times shorter than the integration time for IQ demodulation (see \ref{sec:Phase-space Measurement}). This means that we can safely assume a uniform distribution of $\phiJ$ when the JVS is turned off. 

In that case the $v$ term (see Eq.~\eqref{eq:vdegenerate}) in Eq.~\eqref{eq:gain} averages out and we get 

\begin{equation}
\langle \aout \rangle = u \langle \ain \rangle 
\end{equation}
exactly as in the nondegenerate case. 

The output noise is insenstive to this phase noise. For inputs in the thermal ground state we get 
\begin{equation}
\langle \aout^\dagger \aout \rangle = |u^2| \langle \ain^\dagger \ain \rangle + |v^2| \langle \ain \ain^\dagger \rangle = |v|^2 = |u|^2-1,
\end{equation}
exactly as in the nondegenerate case. This explains why with the JVS turned of in Fig.~\ref{fig:fig2}(a), no particular feature is observed at exact degeneracy $\fdc = 2\fsignal$.

\subsection{\label{sec:Simulation} Effect of JVS RF leakage on the ICTA}

To investigate the effect of leakage from the JVS RF tone on the ICTA, we use a custom semiclassical simulation code (see supplementary information of \cite{nehra2026broadband}).

\begin{figure}[t]
	\centering
	\includegraphics[width = 0.8\textwidth]{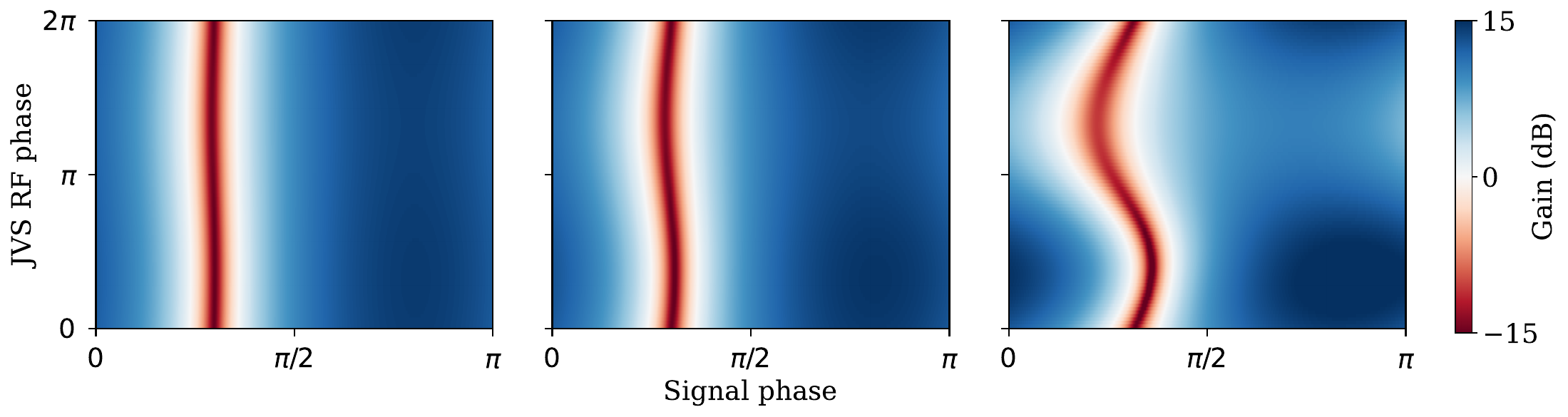}
	\caption{\label{fig:figS6}
			\textbf{Effect of RF leakage on phase-sensitive gain.}
			\textbf{(a--c)} Simulated degenerate gain as a function of input signal phase and leaked RF-tone phase at fixed Josephson frequency phase for three different leakage powers $P_\mathrm{leak}$ at fixed Josephson frequency $\fJVS=\fdc = \SI{12}{\giga\hertz}$. 
			For weak leakage, $P_{\mathrm{leak}}=\SI{-120}{dBm}$, the gain remains essentially unchanged, indicating that the leaked tone has negligible influence on the phase-sensitive response. 
			At $P_{\mathrm{leak}}=\SI{-110}{dBm}$, a weak phase modulation of the gain appears due to interference between Josephson frequency and leaked RF tone. 
			For stronger leakage, $P_{\mathrm{leak}}=\SI{-95}{dBm}$, the response develops a pronounced modulation of both magnitude and phase of the gain. These results explain the experimentally observed gain modulation in Fig.~\ref{fig:fig4}. 
			}
\end{figure}

To model RF leakage from the JVS, we introduce an additional RF port coupled to the ICTA and simulate the gain as a function of both the input-signal phase and the leaked RF-tone phase with fixed Josephson frequency phase, using parameters matching the experiments in main text. Fig.~\ref{fig:figS6}(a) shows the response for a leaked RF power of \SI{-120}{dBm} at \SI{12}{\giga\hertz}, where no visible effect on the degenerate gain is observed which corresponds to the experimental measurement in Fig.~\ref{fig:fig3}(b,d). Increasing the leakage power to \SI{-110}{dBm}, shown in Fig.~\ref{fig:figS6}(b), produces a weak modulation arising from constructive and destructive interference between the leaked JVS RF tone and Josephson frequency, consistent with the measurement shown in Figs.~\ref{fig:fig4}(a,c). At the largest leakage power, \SI{-95}{dBm}, corresponding to Fig.~\ref{fig:figS6}(c), a clear nonlinear modulation of both phase-dependence and magnitude of gain appears. A similar behavior is observed experimentally in Fig.~\ref{fig:fig4}(b, d). These results confirm that the observed gain modulations can be explained by RF leakage interfering with the phase reference transmitted via the order parameter. In consequence, at low RF power, where leaked power becomes neglibile, the phase reference for the ICTA must be transmitted via the order parameter, i.e.\ via phase stabilization of the dc bias provided by the JVS, rather than RF leakage.

\subsection{\label{sec:Saturation in the degenerate regime} Saturation in the degenerate regime}

\begin{figure}[H]
	\centering
	\includegraphics[width = 1\textwidth]{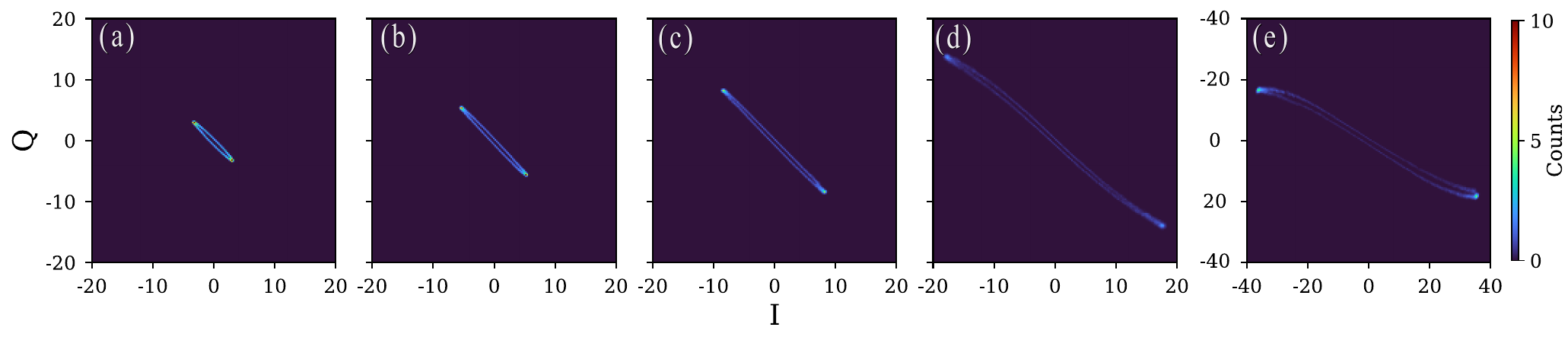}
	\caption{\label{fig:figS7}
		\textbf{Phase-space evolution with increasing ICTA gain.} 
		\textbf{(a)--(e)} Phase-space distributions for five distinct nonlinearity regimes, tuned by varying the SQUID critical current \(I_C\) while maintaining constant JVS stabilization. The resulting gain values are 10, 14, 21, 27, and \SI{31}{\decibel}, corresponding to  \(I_C = 55, 60, 70, 80, 90~\mathrm{nA}\), respectively (Note different sacle on panel (e)).
		To enhance the clarity of the figure, we average over 10 histogram samples, significantly reducing the noise compared to the single-shot data in Fig.~\ref{fig:fig5}.
	}
\end{figure}

In Fig.~\ref{fig:figS7}, we present phase-space gain measurements for five different gains with all other parameters fixed. 
Figs.~\ref{fig:figS7}(a--e) correspond to gain values of 10, 14, 21, 27 and \SI{31}{\decibel}, or   $I_C = 55, 60, 70, 80, \SI{90}{\nano\ampere}$, respectively. While squeezing remains robust and elliptical for gains up to \(27~\mathrm{dB}\), higher nonlinearities due to saturation of the ICTA at \SI{31}{\decibel} cause the ellipse to deform into an S-shape. At the same time, squeezing along the narrow axis of the deformed ellipse is markedly degraded.

\end{document}